\documentclass[11pt,a4papper]{article}

\usepackage{jcappub}
\usepackage{amsmath}
\usepackage{amssymb}
\usepackage{braket}
\usepackage{graphicx}
\usepackage{hyperref}
\usepackage{mathrsfs}
\usepackage{subcaption}
\usepackage{empheq}
\usepackage{appendix}
\usepackage{natbib}
\usepackage[dvipsnames]{xcolor}
\usepackage[top=1.2in, bottom=0.6in,left=0.8in, right=0.8in,includefoot]{geometry}       
\title{Frame-invariant approach to \\higher-dimensional scalar-tensor gravity}

\author{Alexandros Karam,}
\author{Angelos Lykkas,}
\author{Kyriakos Tamvakis}

\affiliation{Department of Physics, Section of Theoretical Physics, University of Ioannina,\\   GR-451 10, Ioannina, Greece}

\emailAdd{alkaram@cc.uoi.gr}
\emailAdd{alykkas@cc.uoi.gr}
\emailAdd{tamvakis@uoi.gr}

\abstract{The recent interest in modified theories of gravity, involving some type of non-minimal coupling to the Ricci scalar, and the calculation of cosmological observables in the Einstein or the Jordan frame, motivate the formulation of these theories in terms of quantities that are invariant under frame transformations. Furthermore, in view of the description of gravity and its geometry motivated by string theory, such a formulation could be extended to include theories of extra spatial dimensions. In the present article, we generalize the construction of frame-invariant quantities, concerning a general, $D$-dimensional scalar-tensor theory. Then, we limit our scope to the $5D$ braneworld scenario, where we study thick brane solutions that are localized on the $3$-brane and extend the invariant formulation to the case of multiple scalar fields (non-)minimally coupled to gravity.}
\numberwithin{equation}{section}

\definecolor{orange}{rgb}{0.9,0.2,0}
\definecolor{brown}{rgb}{0.7,0.3,0.2}
\definecolor{fuxia}{rgb}{1,0,1}
\definecolor{skyblue}{rgb}{0,0.1,0.9}
\definecolor{violetred}{rgb}{0.8,0.13,0.56}
\definecolor{deeppink}{rgb}{1.00,0.08,0.5}
\definecolor{pink}{rgb}{1.00,0.75,0.80}
\definecolor{orchid}{rgb}{0.85,0.44,0.84}
\definecolor{lightpink}{rgb}{1.00,0.71,0.76}
\definecolor{bluish}{rgb}{0,0.6,0.8}
\begin{document}

\maketitle
%%%%%%%%%%%%%%%%%%%%%%%%%%%%%%%%%%%%%%
%%%%%%%%%%%%%%%%%%%%%%%%%%%%%%%%%%%%
\section{Introduction}

The extension of General Relativity into a scalar-tensor theory of gravitation has been a topic of much interest, dating back to the Brans-Dicke theory~\cite{Brans1961}. Relatively recent efforts towards higher dimensional theories lead to extra scalar fields coupled non-minimally to gravity. A common feature of these models is a coupling term $f(\phi)R$ between the scalar field $\phi$ and the Ricci scalar, while the scalar field is absent from the matter action. This is known as the \emph{Jordan frame} and is beneficial for studying physical properties of the theory, as for example masses, coupling constants, decay rates, cross sections, etc. Nevertheless, the equations governing the gravitational dynamics of the theory are complicated and difficult to interpret physically. A rescaling of the metric, followed by a reparametrization of the scalar field can transform the theory into different \emph{conformal frames} or even into the standard Einstein-Hilbert form. This is has been dubbed the \emph{Einstein frame}, where the matter sector gains a factor -- dependent on the non-minimal coupling -- due to the metric rescaling. Common wisdom on the subject accepts that formulations in different conformal frames are equivalent at the classical level~\cite{Capozziello1997,Dick1998,Faraoni1999,Faraoni1999a,Flanagan2004,Bhadra2007,Nozari2009,Capozziello2010,
Corda2011,Quiros2013,Chiba2013,Postma2014,Domenech2015,Bahamonde2016,Bhattacharya2017,Bahamonde2017,
Azri2018,Rashidi2017,Shokri2017,Hyun2017,Bhattacharya2018,Odintsov2018}, although that does not guarantee physical equivalence when quantum corrections  are considered~\cite{Nojiri2001,Kamenshchik2015,Herrero-Valea2016,Pandey2016,Pandey2017,Ohta2017,Ruf2018}. Adopting the point of view that formulations in different conformal frames are classically equivalent it is useful to define quantities that are invariant under frame transformations and formulate the scalar-tensor theory in terms of them. Such a formulation has been introduced in Refs.\cite{Jaerv2015a,Jaerv2015} and has been employed in the analysis and predictions of various models~\cite{Kuusk2016,Kuusk2016a,Jaerv2017,Karam2017,Hohmann2017,Hohmann2018}.

In what follows we start with a general scalar-tensor action describing a theory of gravitation in arbitrary $D$ dimensions, non-minimally coupled to a real scalar field
\begin{equation*}\label{act}
\mathcal{S}=\int\!\mathrm{d}^Dx\,\sqrt{-g}\,\left\{\frac{1}{2}{\mathcal{A}}(\Phi)R-\frac{1}{2}{\mathcal{B}}(\Phi)\left(\nabla\Phi\right)^2-\mathcal{V}(\Phi)\right\}+\mathcal{S}_m[e^{2\sigma(\Phi)}g_{\mu\nu},\psi]\,.
\end{equation*}
Each particular model is defined by a set of \emph{model functions} $\left\{{\mathcal{A}},\,{\mathcal{B}},\,{\mathcal{V}},\,\sigma\right\}$ in a conformal frame. It will be shown that a conformal transformation accompanied by a scalar field redefinition preserves the functional form of the above action through a redefinition of the model functions. Then, we can define a set of \emph{frame-invariant} quantities which do not change under the aforementioned conformal transformation and field redefinition. Nevertheless, these quantities depend on the model, i.e. the model functions in the particular frame in which the model is defined. Thus, we proceed to formulate the theory in terms of these invariants and write the action exclusively in terms of them and the chosen metric. 

As an application of this formalism we consider theories of gravity with extra spatial dimensions. Higher-dimensional models with infinite extra dimensions require the localization of gravitational degrees of freedom on a four-dimensional brane~\cite{Rubakov1983,Antoniadis1998,Arkani-Hamed1999,Randall1999,Randall1999a,DeWolfe2000,Gremm2000,Kehagias2001,Guerrero2002}, since gravity permeates throughout the bulk. Such a brane could be dynamically localized in the higher-dimensional continuum. Additionally, the Standard Model particles are confined on the $3$-brane. This scheme has been dubbed the \emph{braneworld scenario}. A number of models have been proposed~\cite{Bogdanos2006,Bogdanos2007,Farakos2007,Setare2009,Dzhunushaliev2010,Liu2011,Wudka2011,Herrera-Aguilar2012,Yang2012,Liu2012,Ahmed2013,Bazeia2014,Zhong2014,Bazeia2015,Bazeia2015a,Bazeia2015b,
Zhong2016,Chakraborty2016} realizing the idea of modeling the brane with a scalar field configuration and the gravitational part of the action is non-minimally coupled to a fundamental or an auxiliary scalar field, as in the case of $f(R)$ gravity. In the present  article, we examine $5D$ models of scalars coupled non-minimally to gravity, in the framework of a frame-invariant formulation and focus on the study of thick braneworld solutions. 

The paper is organized as follows: In section 2 we generalize the frame-invariant formalism to $D$ dimensions and derive the equations of motion arising from a general action depending only on these invariant quantities. We also discuss the frame-invariant formulation of $f(R)$ theories in their scalar representation. In section 3 we introduce a warped ansatz for the $5D$ metric in terms of frame-invariant quantities and proceed to give examples of thick braneworld solutions. In section 4 we study the localization of gravity on the brane. In section 5 we discuss the frame-invariant formulation of multiscalar theories. Finally, in the last section we state our conclusions.

%%%%%%%%%%%%%%%%%%%%%%%%%%%%%%%%%%%%%%%%%%
%%%%%%%%%%%%%%%%%%%%%%%%%%%%%%%%%%%%%%%%%%%
\section{The Frame-Invariant Formalism}

The most general $D$-dimensional scalar-tensor action has the form~\cite{Flanagan2004}
\begin{equation}\label{act1}
\mathcal{S}=\int\!\mathrm{d}^Dx\,\sqrt{-g}\,\left\{\frac{1}{2}{\mathcal{A}}(\Phi)R-\frac{1}{2}{\mathcal{B}}(\Phi)\left(\nabla\Phi\right)^2-\mathcal{V}(\Phi)\right\}+\mathcal{S}_m[e^{2\sigma(\Phi)}g_{\mu\nu},\psi]\,.
\end{equation}
In this section we shall limit our analysis in the simplest case of one real scalar field $\Phi$, although the multiple scalar fields case can be analyzed in an analogous manner (see section \ref{Nscalar}). In this work we have assumed units where $c\!=\!1\!=\!M_*^{D-2}$, with $M_*$ being the effective Planck mass. The model functions ${\mathcal{A}},\,{\mathcal{B}},\,\mathcal{V}$ and $\sigma$ are unspecified dimensionless functions of the scalar field $\Phi$, depending on each particular model. Here, $\mathcal{A}$ is the non-minimal coupling between the scalar field and the Ricci scalar, $\mathcal{B}$ is the non-minimal kinetic function, $\mathcal{V}$ is the scalar potential and the matter action is denoted by $\mathcal{S}_m$ containing additional matter fields represented by $\psi$ and can be taken to be the action of the Standard Model. Finally, $\sigma$ denotes the coupling of the scalar field to the matter action $\mathcal{S}_m$ and is often called the \emph{matter coupling}. 

 Consider now a general \emph{Weyl rescaling} of the metric\footnote{In what follows we have assumed that $\bar{\gamma}$ is a smooth function -- at least up to order $2$ -- in order to avoid introducing singularities through a conformal transformation.}~\cite{Flanagan2004,Jaerv2015}
\begin{equation}\label{resc1}
g_{\mu\nu}\,=\,e^{2\bar{\gamma}(\bar{\Phi})}\,\bar{g}_{\mu\nu}\,,
\end{equation}
accompanied by a redefinition of the scalar field
\begin{equation}\label{resc2}
\Phi\,=\,\bar{f}(\bar{\Phi})\,.
\end{equation}
The resulting form of the action -- up to boundary terms -- retains its functional form 
\begin{equation}
\mathcal{S}=\int\!\mathrm{d}^Dx\,\sqrt{-\bar{g}}\left\{\frac{1}{2}\bar{\mathcal{A}}(\bar{\Phi})\bar{R}-\frac{1}{2}\bar{\mathcal{B}}(\bar{\Phi})\left(\bar{\nabla}\bar{\Phi}\right)^2-\bar{\mathcal{V}}(\bar{\Phi})\right\}+\mathcal{S}_m[e^{2\bar{\sigma}(\bar{\Phi})}\bar{g}_{\mu\nu},\psi]\,,
\end{equation}
provided that
\begin{align}\label{barred1}
\bar{\mathcal{A}}(\bar{\Phi})&=e^{(D-2)\bar{\gamma}}\mathcal{A}(\Phi)\,,\\
\bar{\mathcal{B}}(\bar{\Phi})&=e^{(D-2)\bar{\gamma}}\left\{\left(\bar{f}'\right)^2{\mathcal{B}}-(D-1)(D-2)\left(\bar{\gamma}'\right)^2{\mathcal{A}}-2(D-1)\bar{\gamma}'{\mathcal{A}}'\bar{f}'\right\}\,,\\
\bar{\mathcal{V}}(\bar{\Phi})&=e^{D\bar{\gamma}}\,{\mathcal{V}}(\Phi)\,,\\
\bar{\sigma}(\bar{\Phi})&=\sigma(\Phi)+\bar{\gamma}(\bar{\Phi})\,,
\end{align}
where the unbarred quantities are meant as functions of $\Phi$, while the barred ones as functions of $\bar{\Phi}$. The primes denote differentiation with respect to the corresponding argument, i.e. $\bar{\gamma}'\!=\!\mathrm{d}\bar{\gamma}/\mathrm{d}\bar{\Phi}$, while ${\mathcal{A}}'\!=\!\mathrm{d}\mathcal{A}/\mathrm{d}\Phi$.
Using the transformations \eqref{resc1} and \eqref{resc2} we can fix two out of the four independent model functions. For example, a standard parametrization corresponds to ${\mathcal{B}}\!=\!1,\,\sigma\!=\!0$, while ${\mathcal{A}}(\Phi),\,{\mathcal{V}}(\Phi)$ are independent -- known as the Jordan frame in Boisseau, Esposito-Far{\'e}se, Polarski parametrization~\cite{Boisseau2000,Esposito-Farese2001}. Alternatively, the Jordan frame in Brans-Dicke-Bergmann-Wagoner parametrization~\cite{Bergmann1968,Wagoner1970} is ${\mathcal{A}}=\Phi,\,{\mathcal{B}}=\omega(\Phi)/\Phi,\,\sigma=0,\,{\mathcal{V}}(\Phi)$.

Next, we introduce the following quantities which are invariant under a change of frame\footnote{Alternatively, one can define $\mathcal{I}_2$ as $\mathcal{I}_2^{D-2}$, which is also invariant under the transformations \eqref{resc1}, \eqref{resc2}.}
\begin{align}
\mathcal{I}_1&\equiv\frac{e^{(D-2)\sigma(\Phi)}}{\mathcal{A}(\Phi)}\,,\label{i1}\\
{\mathcal{I}}_2&\equiv\frac{{\mathcal{V}}(\Phi)}{\left({\mathcal{A}}(\Phi)\right)^{D/(D-2)}}\,,\label{i2}\\
{\mathcal{I}}_3&\equiv\pm\int\!\mathrm{d}\Phi\,\sqrt{{\mathcal{F}}(\Phi)}\,,\label{i3}
\end{align}
where the quantity $\mathcal{F}$ is defined as
\begin{equation}
{\mathcal{F}}\equiv\frac{1}{2}\frac{\mathcal{B}}{\mathcal{A}}+\frac{1}{2}\frac{(D-1)}{(D-2)}\left(\frac{\mathcal{A}'}{\mathcal{A}}\right)^2\,.
\end{equation}
Note that this quantity rescales as
\begin{equation}
\bar{\mathcal{F}}=\left(\bar{f}'\right)^2\,\mathcal{F}\,.
\end{equation}
The requirement $\mathcal{F}>0$ is directly related to the absence of ghosts. 

The first invariant quantity $\mathcal{I}_1$ characterizes the type of non-minimal coupling of the theory. Clearly, if $\mathcal{I}_1$ is constant, the theory is minimally coupled. The second invariant $\mathcal{I}_2$ encompasses the self-interacting dynamics of the scalar field and represents the invariant potential. The invariant $\mathcal{I}_3$ can be seen as the volume integral over the field space of the scalar field\footnote{This can be better understood in the multiscalar formulation of the theory, where the measurement in \eqref{i3} is promoted to $\mathrm{d}\Phi^1\wedge\mathrm{d}\Phi^2\wedge\ldots\wedge\mathrm{d}\Phi^n$, for $n$ scalar fields.} and therefore corresponds to the invariant distance in that space~\cite{Jaerv2017}. If phantom fields are considered in the theory, one has to recover the minus sign in \eqref{i3}. In what follows we consider only the positive branch in \eqref{i3}.

Next, we may introduce an \emph{invariant metric}~\cite{Jaerv2015} as
\begin{equation}
\hat{g}_{\mu\nu}\equiv{\mathcal{A}}^{2/(D-2)}g_{\mu\nu}\,.
\end{equation}
Note that this metric is not unique in the sense that it can be multiplied with any combination of the invariant quantities ${\mathcal{I}}_i$. For example, another definition of an invariant metric could be $\breve{g}_{\mu\nu}\,=\,\left({\mathcal{I}}_i{\mathcal{A}}\right)^{2/(D-2)}\,g_{\mu\nu}$.

The resulting \emph{``frame-invariant"} form of the action is -- up to surface terms -- given by~\cite{Jaerv2015}
\begin{equation}
\mathcal{S}=\int\!\mathrm{d}^Dx\,\sqrt{-\hat{g}}\left\{\frac{1}{2}\hat{R}-\left(\hat{\nabla}\mathcal{I}_3\right)^2-\mathcal{I}_2\right\}+\mathcal{S}_m[\mathcal{I}_1^{2/(D-2)}\hat{g}_{\mu\nu},\psi]
\end{equation}
We are using the shorthand notation $(\hat{\nabla}\mathcal{I}_i)^2=\hat{g}^{\mu\nu}\nabla_\mu\mathcal{I}_i\nabla_\nu\mathcal{I}_i$ and do not include the hats in the covariant derivatives. It is understood that the covariant derivatives are acting on the fields with respect to the spacetime coordinates of the corresponding metric.\footnote{It will become clear by the introduction of another invariant metric later on, that if we were interested in comparing results between ``frames'' we would have to also specify the spacetime coordinates.} Note that the action is expressed in a way reminiscent of the Einstein frame. Variation with respect to $\hat{g}_{\mu\nu}$ gives the Einstein equation
\begin{equation}\label{eq1}
\hat{G}_{\mu\nu}-2\left(\nabla_{\mu}{\mathcal{I}}_3\right)\left(\nabla_{\nu}{\mathcal{I}}_3\right)+\hat{g}_{\mu\nu}\left(\hat{\nabla}{\mathcal{I}}_3\right)^2+\hat{g}_{\mu\nu}{\mathcal{I}}_2=2\hat{T}_{\mu\nu}^{(m)}\,.
\end{equation}
Similarly, variation with respect to the scalar field invariant ${\mathcal{I}}_3$ gives the equation of motion
\begin{equation}\label{eq2}
\frac{1}{\sqrt{-\hat{g}}}\nabla_{\mu}\left(\sqrt{-\hat{g}}\,\hat{g}^{\mu\nu}\nabla_{\nu}{\mathcal{I}}_3\right)=\hat{\Box}\mathcal{I}_3=\frac{1}{2}\frac{\mathrm{d}\mathcal{I}_2}{\mathrm{d}\mathcal{I}_3}-\frac{\hat{T}}{4{\mathcal{I}}_1}\frac{\mathrm{d}\mathcal{I}_1}{\mathrm{d}\mathcal{I}_3}\,,
\end{equation}
where $\hat{T}=\hat{g}^{\mu\nu}\hat{T}_{\mu\nu}^{(m)}$. 

A particular alternative metric is the one for which the non-minimally coupled scalar does not enter explicitly into the matter action through the metric. It is defined as
\begin{equation}\label{invmetric1}
\tilde{g}_{\mu\nu}\equiv e^{2\sigma}g_{\mu\nu}
\end{equation}
and is related to the invariant metric $\hat{g}_{\mu\nu}$ as
\begin{equation}\label{invmetric2}
\tilde{g}_{\mu\nu}={\mathcal{I}}_1^{2/(D-2)}\hat{g}_{\mu\nu}\,.
\end{equation}
Thus, it is manifestly invariant.  The action in terms of this metric is
\begin{equation}\label{invact}
\mathcal{S}=\int\!\mathrm{d}^Dx\,\sqrt{-\tilde{g}}\left\{\frac{1}{2{\mathcal{I}}_1}\tilde{R}-\frac{1}{{\mathcal{I}}_1}\left(\tilde{\nabla}{\mathcal{I}}_3\right)^2+\frac{(D-1)}{2(D-2)}{\mathcal{I}}_1^{-3}\left(\tilde{\nabla}{\mathcal{I}}_1\right)^2-{\mathcal{I}}_1^{-D/(D-2)}{\mathcal{I}}_2\right\} +\mathcal{S}_m[\tilde{g}_{\mu\nu},\psi]\,,
\end{equation}
which we refer to as the \emph{invariant Jordan frame action}. The resulting equations of motion are
\begin{align}\label{eq3}
\tilde{G}_{\mu\nu}&=\mathcal{I}_1\,\tilde{T}_{\mu\nu}-\tilde{g}_{\mu\nu}\left[\left(\tilde{\nabla}\mathcal{I}_3\right)^2-\frac{\mathcal{I}^{-2}_1}{2}\,\frac{D-1}{D-2}\left(\tilde{\nabla}\mathcal{I}_1\right)^2+\mathcal{I}_1\mathcal{I}_4\right]+2\,\nabla_\mu\mathcal{I}_3\nabla_\nu\mathcal{I}_3-\nonumber\\
&\qquad-\frac{D-1}{D-2}\mathcal{I}_1^{-2}\,\nabla_\mu\mathcal{I}_1\nabla_\nu\mathcal{I}_1+\mathcal{I}_1\nabla_\mu\nabla_\nu\mathcal{I}_1^{-1}-\mathcal{I}_1\,\tilde{g}_{\mu\nu}\,\tilde{\Box}\,\mathcal{I}_1^{-1}\,,
\end{align}
\begin{align}\label{eq4}
&\tilde{\Box}\mathcal{I}_3-\frac{1}{\mathcal{I}_1}\left(\tilde{\nabla}\mathcal{I}_1\right)\cdot\left(\tilde{\nabla}\mathcal{I}_3\right)+\left(\tilde{\nabla}\mathcal{I}_3\right)^2+\nonumber\\&+\frac{\mathrm{d}\mathcal{I}_1}{\mathrm{d}\mathcal{I}_3}\left\{\frac{3}{4\mathcal{I}_1^3}\,\frac{D-1}{D-2}\left(\tilde{\nabla}\mathcal{I}_1\right)^2-\frac{1}{2\mathcal{I}_1^2}\,\frac{D-1}{D-2}\,\tilde{\Box}\mathcal{I}_1+\frac{1}{2\mathcal{I}_1}\left(\tilde{\nabla}\mathcal{I}_3\right)^2-\frac{1}{4\mathcal{I}_1}\tilde{R}\right\}=\frac{\mathcal{I}_1}{2}\,\frac{\mathrm{d}\mathcal{I}_4}{\mathrm{d}\mathcal{I}_3}\,,
\end{align}
where ${\mathcal{I}}_4\equiv{\mathcal{I}}_2/{\mathcal{I}}_1^{D/(D-2)}$.

Closing this section we also consider the case of $f(R)$ gravity for which there is also a scalar representation in terms of a scalar field non-minimally coupled to the Ricci scalar (for some reviews see Refs.\cite{Sotiriou2008,Felice2010,Nojiri2011,Capozziello2011,Nojiri2017}). The general $D$-dimensional action of $f(R)$ gravity is
\begin{equation}
\mathcal{S}=\int\!\mathrm{d}^Dx\,\sqrt{-g}\,f(R)+{\mathcal{S}}_m[e^{2\sigma}g_{\mu\nu},\psi]\,.
\end{equation}
The corresponding scalar representation is (for $f''(R)\!\neq\!0$)
\begin{equation}\label{scalrep}
\mathcal{S}=\int\!\mathrm{d}^Dx\,\sqrt{-g}\left\{F'(\chi)R-{\mathcal{V}}(\chi)\right\}+{\mathcal{S}}_m[e^{2\sigma}g_{\mu\nu},\psi]\,,
\end{equation}
where
\begin{equation}
{\mathcal{V}}(\chi)\,=\,\chi\,F'(\chi)\,-F(\chi)\,,
\end{equation}
and $\chi$ is an auxiliary scalar field, satisfying the on-shell condition $\chi\!=\!R$.
This action is a particular case of \eqref{act1} for which, in this frame, we have the relations
\begin{equation}
{\mathcal{A}}(\chi)=2\,F'(\chi),\qquad{\mathcal{B}}(\chi)=0\quad{\text{and}}\quad{\mathcal{V}}(\chi)=\chi F'(\chi)-F(\chi)\,.
\end{equation}
Introducing the invariants in the same way as in the case of a fundamental scalar field we may ultimately express the action in terms of them. Nevertheless, now there is an extra relation between the invariants $\mathcal{I}_1$ and $\mathcal{I}_3$. We will revisit this in the multifield case in section \ref{Nscalar}.

%%%%%%%%%%%%%%%%%%%%%%%%%%%%%%%%%%%%%%%%%%%%%%%%%%%%%%%%%%%%%%%%%%%%%%%%%%%%%%%%%%%%%%%%%%%%%%%%%%%%
\section{Thick Branes}
In this section we shall briefly review the paradigm of \emph{thick branes}\footnote{In the context of braneworld scenarios, domain walls are categorized as thin or thick with respect to the surface where matter is localized. For example, it has been well studied~\cite{Randall1999,Randall1999a,DeWolfe2000} that thin branes are described by a $\delta$ function potential corresponding to an ideal surface with matter fields on it.} before proceeding to their frame-invariant formulation.

Consider a $D=5$ theory based on \eqref{act1} with ${\mathcal{B}}=1$, while ignoring ${\mathcal{S}}_m$ and setting $\sigma=0$. The equations of motion are
\begin{equation}
{\mathcal{A}}\left(R_{MN}-\frac{1}{2}g_{MN}R\right)-\nabla_M\nabla_N{\mathcal{A}}+g_{MN}\Box{\mathcal{A}}=\nabla_M\Phi\nabla_N\Phi-g_{MN}\left(\frac{1}{2}\left(\nabla\Phi\right)^2+{\mathcal{V}}\right)\,,
\end{equation}
\begin{equation}
\Box\Phi-{\mathcal{V}}'+\frac{1}{2}R{\mathcal{A}}'=0\,.
\end{equation}
We introduce a warped ansatz for the metric in terms of a \emph{warp function} $Z$ as
\begin{equation}
g_{MN}\,=\,\left(\begin{array}{cc}
e^{2Z(y)}\,\eta_{ \mu\nu}\,&\,0\\
\,&\,\\
0\,&\,1\,
\end{array}\right)\,,
\end{equation}
where we have denoted with $\mu,\,\nu$ the standard four-dimensional spacetime indices, while $y$ stands for the fifth spatial coordinate and $M,N$ refer to the five-dimensional coordinate indices. For the four-dimensional Minkowski metric we adopt the sign convention $\eta_{\mu\nu}\!\in\!\mathbb{R}^{1,3}$. Note that the extra dimension is assumed to be flat and infinite as well as respecting no \textit{a priori} symmetry. We have explicitly assumed that the warp function depends only on the fifth coordinate and is a smooth function of it. For a scalar field  $\Phi(y)$ depending only on the fifth coordinate the equations of motion take the form
\begin{equation}\label{reveq1}
\left(\dot{\Phi}\right)^2=-\ddot{\mathcal{A}}-3{\mathcal{A}}\ddot{Z}+\dot{Z}\dot{\mathcal{A}}\,,
\end{equation}
\begin{equation}\label{reveq2}
2\,{\mathcal{V}}=-\ddot{\mathcal{A}}-3{\mathcal{A}}\left(\ddot{Z}+4(\dot{Z})^2\right)-7\dot{Z}\dot{\mathcal{A}}\,,
\end{equation}
where the dots refer to differentiation with respect to the fifth coordinate. Next, we may choose a quadratic form for the non-minimal coupling function
\begin{equation}
{\mathcal{A}}(\Phi)\,=\,1-\frac{\alpha}{2}\Phi^2\,,
\end{equation}
with $\alpha\!>\!0$ and dimensionless. Inserting a particular kink configuration for the scalar field
\begin{equation}\label{bounce}
\Phi(y)=\upsilon\tanh(by)
\end{equation}
and considering the above system of \eqref{reveq1}-\eqref{reveq2} as equations determining the warp function and the supporting potential in terms of this configuration, we obtain a solution for the warp function\footnote{Without loss of generality we have assumed the following initial conditions: $\dot{Z}(0)=0=Z(0)$.}
\begin{equation}\label{warp}
Z=-\lambda\ln(\cosh(by) )=\frac{\lambda}{2}\ln\left(1-\frac{\Phi^2}{\upsilon^2}\right)\,,
\end{equation}
for the parameter values
\begin{equation}
\lambda=2\,\alpha^{-1}-6,\qquad \upsilon^2=3\lambda/(1-\alpha)\,,
\end{equation}
valid for $0<\alpha<1/3$. This stems from the constraint that the warp function $e^{2Z}$ has to vanish at distances far away from the brane, $y\rightarrow\pm\infty$, leading to the condition that $\lambda\!>\!0$. The corresponding supporting scalar potential, expressed in terms of the field $\Phi$ turns out to be a quartic positive definite potential. An analogous solution can be obtained also for the same configuration \eqref{bounce} even for ${\mathcal{A}}=1$, namely~\cite{Kehagias2001}
\begin{equation}\label{warpKT}
Z=-\frac{\upsilon^2}{9}\ln(\cosh(by))-\frac{\upsilon^2}{36}\tanh^2(by)=\frac{\upsilon^2}{18}\ln\left(1-\frac{\Phi^2}{\upsilon^2}\right)-\frac{1}{36}\Phi^2\,.
\end{equation}
Solutions of the form of \eqref{warp} and \eqref{warpKT} exhibit a $\mathbb{Z}_2$ symmetry with respect to the $y$ coordinate, leading to an asymptotic $AdS_5/\mathbb{Z}_2$ bulk geometry. It has been well documented that the above warp functions and the associated geometry lead to graviton localization~\cite{Kehagias2001,Liu2012}.

In order to investigate a frame-invariant generalization of the above configurations we may start from the action \eqref{invact}, which, for $D=5$ in terms of the metric $\tilde{g}_{MN}$, takes the form
\begin{equation}\label{invact2}
{\mathcal{S}}=\int\!\mathrm{d}^5x\,\sqrt{-\tilde{g}}\left\{\frac{1}{2{\mathcal{I}}_1}\tilde{R}-\frac{1}{{\mathcal{I}}_1}\left(\tilde{\nabla}{\mathcal{I}}_3\right)^2-\frac{2}{3}{\mathcal{I}}_1^{-3}\left(\tilde{\nabla}{\mathcal{I}}_1\right)^2-{\mathcal{I}}_1^{-5/3}{\mathcal{I}}_2\right\}+\mathcal{S}_m[\tilde{g}_{\mu\nu},\,\chi]\,.
\end{equation}
Next, we consider the metric ansatz
\begin{equation}\label{ans}
\tilde{g}_{MN}={\mathcal{I}}_1^{2/3}\begin{pmatrix}
e^{2\mathcal{Z}}\eta_{\mu\nu}&0\\0&1
\end{pmatrix}\,,
\end{equation}
where again the warp factor is assumed to depend on the fifth coordinate denoted by $y$. Similarly, the scalar field is only a function of the fifth coordinate as well. Therefore, all the invariant quantities depend only on the fifth coordinate. In terms of this ansatz, the components of the Einstein tensor are
\begin{equation}
\tilde{G}_{\mu\nu}=\eta_{ \mu\nu}\,e^{2{\mathcal{Z}}}\left\{3\ddot{\mathcal{Z}}+6\dot{\mathcal{Z}}^2+\frac{\ddot{\mathcal{I}}_1}{{\mathcal{I}}_1}+3\dot{\mathcal{Z}}\frac{\dot{\mathcal{I}}_1}{{\mathcal{I}}_1}-\frac{2}{3}\left(\frac{\dot{\mathcal{I}}_1}{{\mathcal{I}}_1}\right)^2\right\}\,,\\
\end{equation}
\begin{equation}
\tilde{G}_{55}=6\dot{\mathcal{Z}}^2+\frac{2}{3}\left(\frac{\dot{\mathcal{I}}_1}{{\mathcal{I}}_1}\right)^2+4\dot{\mathcal{Z}}\frac{\dot{\mathcal{I}}_1}{{\mathcal{I}}_1}\,.
\end{equation}
The resulting equations of motion are
\begin{equation}\label{eqs1}
3\ddot{\mathcal{Z}}+2\left(\dot{\mathcal{I}}_3\right)^2=0\,,
\end{equation}
\begin{equation}\label{eqs2}
3\ddot{\mathcal{Z}}+12\dot{\mathcal{Z}}^2+{\mathcal{I}}_2=0\,.
\end{equation}
 
 In what follows, we assume again that the scalar field represented by the invariant ${\mathcal{I}}_3$ will correspond to a given configuration. Then, the warp factor ${\mathcal{Z}}$ will be determined by \eqref{eqs1}, while the potential that can sustain this configuration will be given by \eqref{eqs2}. We may look for a solution where $\dot{\mathcal{I}}_3$ is a function of ${\mathcal{I}}_3$, expandable in powers of ${\mathcal{I}}_3^2$. We assume the simplest choice
 \begin{equation}
 \dot{\mathcal{I}}_3=C_0+C_1{\mathcal{I}}_3^2\,.
 \end{equation}
 Then, we have
 \begin{equation}
 \ddot{\mathcal{Z}}=-\frac{2}{3}\dot{\mathcal{I}}_3^2\quad\Longrightarrow\quad\frac{\mathrm{d}\dot{\mathcal{Z}}}{\mathrm{d}\mathcal{I}_3}=-\frac{2}{3}\dot{\mathcal{I}}_3=-\frac{2}{3}\left(C_0+C_1{\mathcal{I}}_3^2\right)\,,
 \end{equation}
 or, by taking $\dot{\mathcal{Z}}(0)=0={\mathcal{Z}}(0)$,
 \begin{equation}\label{warp1}
 {\mathcal{Z}}=-\frac{2C_0}{9C_1}\ln\left(1+\frac{C_1}{C_0}{\mathcal{I}}_3^2\right)-\frac{1}{9}{\mathcal{I}}_3^2\,.
 \end{equation}
 For $C_1/C_0=-1$ we have
 \begin{equation}
 e^{2{\mathcal{Z}}}=\left(1-{\mathcal{I}}_3^2\right)^{4/9}\,e^{-\frac{2}{9}{\mathcal{I}}_3^2}\,.
 \end{equation}
 Introducing a kink configuration\footnote{An alternative configuration is ${\mathcal{I}}_3=\cosh^{-1}(by)$, leading to
 $$e^{2{\mathcal{Z}}}\,=\,{\mathcal{I}}_3^{2/9}\,e^{-\frac{1}{9}{\mathcal{I}}_3^2}$$
 with an analogous localized behaviour.}
 \begin{equation}
 {\mathcal{I}}_3=\tanh(by)
 \end{equation}
 we obtain the localized warp factor
 \begin{equation}\label{warping}
 e^{2{\mathcal{Z}}}=\left(\cosh(by)\right)^{-8/9}\,e^{-\frac{2}{9}\tanh^2(by)}\,.
 \end{equation}
 This has the localized form of the known Einstein-frame kink solution~\cite{Kehagias2001}. 
 
 The invariant ${\mathcal{I}}_2$ corresponding to the scalar potential in the case of the above example is easily determined from \eqref{eqs2} to be
 \begin{equation}
 {\mathcal{I}}_2=2C_0^2\left(1+\frac{C_1}{C_0}{\mathcal{I}}_3^2\right)^2\left[\,1-\frac{8}{27}{\mathcal{I}}_3^2\left(1+\frac{2}{1+\frac{C_1}{C_0}{\mathcal{I}}_3^2}\right)^2\right]\,.
 \end{equation}
 
 The solution \eqref{warping}, obtained in terms of the frame-invariant quantities ${\mathcal{I}}_3,\,{\mathcal{I}}_1$ and $\tilde{g}_{MN}$, stands also as a solution for the particular case of a theory based on the action
 \begin{equation}
 {\mathcal{S}}=\int\!\mathrm{d}^5x\,\sqrt{-g}\left\{\frac{1}{2}{\mathcal{A}}(\Phi)R-\frac{1}{2}(\nabla\Phi)^2-{\mathcal{V}}(\phi)\right\}\,.
 \end{equation}
 For this theory, the invariant ${\mathcal{I}}_3$ in terms of ${\mathcal{A}}$ is
 \begin{equation}
 {\mathcal{I}}_3=\int\!\mathrm{d}\Phi\left(\frac{3{\mathcal{A}}+4({\mathcal{A}}')^2}{6{\mathcal{A}}^2}\right)^{1/2}\,.
 \end{equation}
 As an example, we may take
 \begin{equation}
 {\mathcal{A}}(\Phi)=1-\frac{\alpha}{2}\Phi^2\quad\Longrightarrow\quad{\mathcal{I}}_3=\frac{1}{\sqrt{2}}\int\!\mathrm{d}\Phi\,\frac{\sqrt{1+\frac{\alpha}{2}\left(\frac{8\alpha}{3}-1\right)\Phi^2}}{1-\frac{\alpha}{2}\Phi^2}\,.
 \end{equation}
 The integration is elementary and can be easily inverted for the conformal choice $\alpha=3/8$. Proceeding for this case we obtain
 \begin{equation}
 {\mathcal{I}}_3=\sqrt{\frac{2}{3}}\,\ln\left(\frac{1+\frac{\sqrt{3}}{4}\Phi}{1-\frac{\sqrt{3}}{4}\Phi}\right)\quad\Longrightarrow\quad\Phi=\frac{4}{\sqrt{3}}\left(\frac{e^{\sqrt{\frac{3}{2}}{\mathcal{I}}_3}-1}{e^{\sqrt{\frac{3}{2}}{\mathcal{I}}_3}+1}\right)=\frac{4}{\sqrt{3}}\tanh\left(\sqrt{\frac{3}{8}}{\mathcal{I}}_3\right)\,.
 \end{equation}
Note that the configuration of $\Phi$ corresponding to our choice above of ${\mathcal{I}}_3=\tanh(by)$ is
\begin{equation}
\Phi=\frac{4}{\sqrt{3}}\,\tanh\left(\sqrt{\frac{3}{8}}\,\tanh(by)\right)\,,
\end{equation}
and has a similar kink-like profile as ${\mathcal{I}}_3$. Therefore, it is evident that the above formulation in terms of the invariant ${\mathcal{I}}_3$ covers a broad class of possible configurations. Using the expression of $\Phi$ as a function of the invariant $\mathcal{I}_3$ one can derive a relation between the $\mathcal{I}_1$ and $\mathcal{I}_3$ invariants, i.e.:
\begin{equation}
\mathcal{I}_1=\cosh^2{\left(\sqrt{\frac{3}{8}}\,\mathcal{I}_3\right)}.
\end{equation}

In the general case ($\alpha<3/8$) the integral gives
\begin{align}
{\mathcal{I}}_3=&\sqrt{1-8\alpha/3}\,\arcsin\left(\sqrt{\frac{\alpha}{2}\left(1-\frac{8\alpha}{3}\right)}\,\Phi\right)-\sqrt{\frac{2\alpha}{3}}\Bigg\{\,\ln\left(\frac{1-\sqrt{\alpha/2}\,\Phi}{1+\sqrt{\alpha/2}\,\Phi}\right)-\nonumber\\
&-\ln\left(\frac{1-(1-8\alpha/3)\sqrt{\alpha/2}\,\Phi+\sqrt{8\alpha/3}\sqrt{1-(1-8\alpha/3)\alpha\Phi^2/2}}{1+(1-8\alpha/3)\sqrt{\alpha/2}\,\Phi+\sqrt{8\alpha/3}\sqrt{1-(1-8\alpha/3)\alpha\Phi^2/2}}\right)\Bigg\}\,.
\end{align}
Note that for a kink-like configuration like $\Phi\sim \tanh y$ or $\Phi\sim\cosh^{-1}y$ the invariant ${\mathcal{I}}_3$ moves between zero and a finite value. Near $\Phi\sim0$ we have ${\mathcal{I}}_3\sim0$.

%%%%%%%%%%%%%%%%%%%%%%%%%%%%%%%%%%%%%%%%%%%%%%%%%%%%%%%%%%%%%%%%%%%%%%%%%%%%%%%%%%%%%%%%%%%%%%%%%%%%%%%%%%%%%%%%%%%%%%%%%%%%%%%%%%%%%%%%%%%%%%%%%%%%%%%%%%%%%%%%%%%%%%%%%%%%%%%%%%%%%%%
\section{Graviton Localization}

The above warped geometries are known to be supported by graviton localization around $y=0$. In order to examine a frame-invariant formulation of this we may consider fluctuations in the $4D$ Minkowski metric according to our ansatz \eqref{ans}
\begin{equation}
\mathrm{d}s^2={\mathcal{I}}_1^{2/3}\,e^{2{\mathcal{Z}}}\left(\eta_{ \mu\nu}+h_{\mu\nu}\right)\mathrm{d}x^\mu\mathrm{d}x^\nu+{\mathcal{I}}_1^{2/3}\,\mathrm{d}y^2\,.
\end{equation}
The fluctuations can be thought of as $\delta\tilde{g}_{MN}=h_{MN}$ where $h_{MN}$ is a symmetric matrix of order $5$ ($D$ in general). Clearly, $h_{MN}$ has $10$ ($2D$) gauge degrees of freedom due to the fact that General Relativity is invariant under spacetime diffeomorphisms. Therefore, after a complete gauge fixing ($h_{5M}=0$, $h^\mu_\mu=0=\partial^\mu h_{\mu\nu}$) we end up with $5$ [or $D(D-3)/2$] physical degrees of freedom for the massless five-dimensional graviton. Assuming that the scalar field $\Phi$ has vanishing fluctuations, $\delta\Phi\!=\!0$, the resulting equations of motion in the transverse-traceless gauge ($h_{\mu}^{\mu}=0=\partial^{\mu}h_{\mu \nu}$) are
\begin{equation}\label{eom1}
\left(e^{-2{\mathcal{Z}}}\,\partial^2+\partial_y^2+4\dot{\mathcal{Z}}\,\partial_y\right)h_{\mu\nu}=0\,,
\end{equation}
where $\partial^2$ refers to the $4D$ coordinates. Replacing $y$ by the coordinate $w=\int\!\mathrm{d}y\,e^{-{\mathcal{Z}}}$, we can rewrite \eqref{eom1} as
\begin{equation}
\left(\partial^2+\partial_w^2+3{\mathcal{Z}}'\,\partial_w\right)h_{\mu\nu}(x,w)=0\,,
\end{equation}
where the prime denotes differentiation with respect to $w$. A solution can be obtained in the form of a plane-wave decomposition $h_{\mu\nu}(x,w)\,=\,e^{-3{\mathcal{Z}}/2}\,\epsilon_{\mu\nu}\,e^{ik\cdot x}\,\psi(w)$ with
$k^2=m^2$ and $\psi(w)$ satisfying the Schr{\"o}dinger-like equation
\begin{equation}\label{local1}
\left(-\frac{\mathrm{d}^2}{\mathrm{d}w^2}+U(w)\right)\psi(w)=m^2\psi(w)\,,
\end{equation}
where
\begin{equation}
U(w)\equiv\frac{3}{2}{\mathcal{Z}}''+\frac{9}{4}{{\mathcal{Z}}'}^2\,.
\end{equation}
Note that \eqref{local1} can be factorized as
\begin{equation}
\left\{\frac{\mathrm{d}}{\mathrm{d}w}+\frac{3}{2}{\mathcal{Z}}'\,\right\}\left\{-\frac{\mathrm{d}}{\mathrm{d}w}+\frac{3}{2}{\mathcal{Z}}'\,\right\}\psi(w)=m^2\,\psi(w)\,.
\end{equation}
This form excludes tachyonic modes and singles out a normalizable zero mode
\begin{equation}
\psi_0(w)=\mathcal{N}e^{3\mathcal{Z}/2}\,.
\end{equation}
For these types of theories, the case of the localization of matter fields on the brane has been studied~\cite{Gremm2000,Pomarol2000,Bajc2000,Oda2000,Kehagias2001,Ghoroku2002,Ghoroku2003,Oda2003,
Koley2005,Melfo2006,Liu2008,Liu2008a,Liu2009,Guerrero2010,Liu2010,Du2013,Guo2013,Sousa2014,
Vaquera-Araujo2015,Jardim2015,Bernardini2016,Zhou2017,Chen2018}. Therefore, we expect a similar result in our case.

%%%%%%%%%%%%%%%%%%%%%%%%%%%%%%%%%%
%%%%%%%%%%%%%%%%%%%%%%%%%%%%%%%%
%%%%%%%%%%%%%%%%%%%%%%%%%%%%%%
%%%%%%%%%%%%%%%%%%%%%%%%%%%%%%%%%
\section{The Multifield Case}\label{Nscalar}

The case of multiple scalar fields can be treated in a similar manner to that of a single field~\cite{Kuusk2016a}. This will prove useful when one considers the case of $f(R)$ gravity coupled minimally to a scalar field $\phi$. Expressing $f(R)$ in the scalar representation, another auxiliary scalar field is introduced in the action. When one formulates the theory in the Einstein frame, the two scalar fields mix non-trivially and the multifield approach treats this case in general, allowing for different types of couplings with gravity and more scalar fields. Let us consider the general action
\begin{equation}
{\mathcal{S}}=\int\!\mathrm{d}^Dx\,\sqrt{-g}\left\{\frac{1}{2}{\mathcal{A}}(\Phi)R-\frac{1}{2}\mathcal{B}_{ab}(\Phi)\left(\nabla\Phi^a\nabla\Phi^b\right)-{\mathcal{V}}(\Phi)\right\}+{\mathcal{S}}_m[e^{2\sigma}g_{\mu\nu},\chi]\,,
\end{equation}
where $\Phi^a$ stands for $N$ scalar fields ($a=1,\,2,\dots,\,N$). In general, the model functions depend on any and all of the scalar fields $\Phi^a$, henceforth referred to as $\Phi$ to alleviate the notation. Note that $\mathcal{B}_{ab}$ is a symmetric matrix and the Einstein summation convention extends to the scalar field indices $a,\,b$, as well. Now, the general field redefinition accompanying a conformal transformation is $\Phi^a=\bar{f}^a(\bar{\Phi})$. The model functions transform as
\begin{align}
\bar{\mathcal{A}}(\bar{\Phi})&=e^{(D-2)\bar{\gamma}}\,{\mathcal{A}}(\Phi)\,,\\
\bar{\mathcal{B}}_{ab}(\bar{\Phi})&=e^{(D-2)\bar{\gamma}}\,\left[\,\bar{f}^{c}_{\,,a}\bar{f}^{d}_{\,,b}{\mathcal{B}}_{cd}-(D-1)(D-2)\bar{\gamma}_{\,,a}\bar{\gamma}_{\,,b}{\mathcal{A}}-2(D-1)\bar{\gamma}_{\left(\,,a\right.}\bar{f}^c_{\left.\,,b\right.)}{\mathcal{A}}_c\right]\,,\\
\bar{\mathcal{V}}(\bar{\Phi})&=e^{D\bar{\gamma}}{\mathcal{V}}\,,\\
\bar{\sigma}(\bar{\Phi})&=\sigma(\Phi)+\bar{\gamma}(\bar{\Phi})\,,
\end{align}
where $\bar{(..)}_{,a}\equiv\partial\bar{(..)}/\partial\bar{\Phi}^a$ and $2\bar{\gamma}_{\left(,a\right.}\bar{f}^c_{\left.\,,b\right.)}\equiv\bar{\gamma}_{,a}\bar{f}^c_{\,,b}+\bar{\gamma}_{,b}\bar{f}^c_{\,,a}$. 

The frame-invariants ${\mathcal{I}}_1,\,{\mathcal{I}}_2$ have the same functional form as in the single field case, i.e.
\begin{equation}
{\mathcal{I}}_1=\frac{e^{(D-2)\sigma}}{{\mathcal{A}}(\Phi)},\qquad{\mathcal{I}}_2=\frac{{\mathcal{V}}(\Phi)}{\left({\mathcal{A}}(\Phi)\right)^{D/(D-2)}}\,,
\end{equation}
while an invariant ${\mathcal{I}}_3$ can be defined in terms of the frame-covariant quantity
\begin{equation}
{\mathcal{F}}_{ab}\,=\,\frac{1}{2}\frac{{\mathcal{B}}_{ab}}{{\mathcal{A}}}\,+\,\frac{1}{2}\frac{(D-1)}{(D-2)}\frac{{\mathcal{A}}_{\,,a}{\mathcal{A}}_{\,,b}}{{\mathcal{A}}^2}\,,
\end{equation}
as
\begin{equation}\label{iotatria}
{\mathcal{I}}_3^{(n)}\equiv\int\sqrt{{\mathcal{F}}_{ab}\,\mathrm{d}\Phi^a\,\mathrm{d}\Phi^b}\,.
\end{equation}
The quantity ${\mathcal{F}}_{ab}$ transforms as $\bar{\mathcal{F}}_{ab}=\bar{f}^c_{\,,a}\bar{f}^c_{\,,b}\,{\mathcal{F}}_{cd}$ under the field redefinition.

Next, we specialize in the $D\!=\!5$ case. The action in terms of the above quantities and the metric
 \begin{equation}
 \tilde{g}_{MN}=e^{2\sigma}g_{MN}
 \end{equation}
reads
\begin{equation}
{\mathcal{S}}=\int\!\mathrm{d}^5x\,\sqrt{-\tilde{g}}\left\{\frac{1}{2{\mathcal{I}}_1}\tilde{R}-\frac{\omega_{ab}}{{\mathcal{I}}_1}\left(\tilde{\nabla}\Phi^a\tilde{\nabla}\Phi^b\right)-{\mathcal{I}}_1^{-5/3}{\mathcal{I}}_2\right\}+{\mathcal{S}_m}[\tilde{g}_{MN},\chi]\,,
\end{equation}
where
\begin{equation}
\omega_{ab}\equiv{\mathcal{F}}_{ab}-\frac{2}{3}\left(\ln{\mathcal{I}}_1\right)_{,a}\left(\ln{\mathcal{I}}_1\right)_{,b}\,.
\end{equation}
If we introduce now the metric ansatz \eqref{ans} and assume dependence of the scalar fields only on the fifth coordinate, the Einstein equations resulting from a variation with respect to $\tilde{g}_{MN}$ are (ignoring the matter action $\mathcal{S}_m$ and setting $\sigma\!=\!0$.)
\begin{equation}
3\ddot{\mathcal{Z}}+2\left({\dot{\mathcal{I}}}_3^{(n)}\right)^2=0\,,
\end{equation}
\begin{equation}
3\ddot{\mathcal{Z}}+12\dot{\mathcal{Z}}^2+2{\mathcal{I}}_2=0\,,
\end{equation}
having the same form as in the case \eqref{eqs1}-\eqref{eqs2} in terms of the multifield invariant ${\mathcal{I}}_3^{(n)}$ defined in \eqref{iotatria}.

%%%%%%%%%%%%%%%%%%%%%%%%%%%%%%%
%%%%%%%%%%%%%%%%%%%%%%%
%%%%%%%%%%%%%%%%%%%%%%%%%%
%%%%%%%%%%%%%%%%%%%%%%%%%%%%
\section{Conclusions}

In the present article we considered a $D$-dimensional theory of gravity coupled to scalar fields through a general action \eqref{act1} and studied its formulation in terms of quantities that are invariant under general Weyl transformations of the metric accompanied by scalar field redefinitions. As a result, only these invariants appeared in the action apart from the metric and the Ricci scalar. We proceeded with the derivation of the equations of motion, expressed exclusively in terms of invariants and the choice of metric, $\hat{g}_{\mu\nu}$ in the case \eqref{eq1}-\eqref{eq2} or $\tilde{g}_{\mu\nu}$ in the case \eqref{eq3}-\eqref{eq4}, in order to study different theories at the level of the equations of motion, instead of the action level. Then, we applied this formalism to the case of $D\!=\!5$ models with an infinite extra dimension and a warped metric of the Randall-Sundrum type, where a thick brane is formed dynamically from the scalar field configuration and its self-interacting potential. We introduced a metric ansatz expressed in terms of frame-invariant quantities and proceeded to find solutions of the equations of motion with the condition that the quantities living in the bulk are localized and well-defined on the brane. Lastly, we examined the case of multiple scalar fields coupled to gravity encompassing the case of $f(R)$ gravity coupled to extra scalar fields.

As was demonstrated, the frame-invariant formalism has proven to be effective in describing the dynamics of modified gravity theories at the level of the equations of motion. Instead of expressing the action of a theory in each frame and carry out the calculations, we can easily analyze it at the level of the equations of motion. The parametrization dependence is shifted to the model dependence of the invariant quantities $\mathcal{I}_i$. This allows us to effortlessly calculate quantities that are dependent on these invariants and make predictions with respect to physical observables in each frame. Moreover, the proposed formalism can be used to readily test and constrain extensions of General Relativity based on a non-minimal coupling.

\bibliographystyle{JHEP}
\bibliography{biblio}

\end{document}